\documentclass{epl}
\usepackage{graphicx}
\usepackage{amsmath,amssymb}

\title{Charge qubit entanglement in double quantum dots}
\author{S. Weiss \email{}\and M. Thorwart \and R. Egger}
\institute{Institut f\"ur Theoretische Physik,
Heinrich-Heine-Universit\"at, D-40225 D\"usseldorf, Germany}


\begin{document}
\maketitle

\begin{abstract}
We study entanglement of charge qubits in a vertical tunnel-coupled 
double quantum dot containing two interacting electrons. 
Exact diagonalization is used to compute the negativity 
characterizing entanglement. 
We find that entanglement can be efficiently generated
and controlled by sidegate voltages, and describe how it
can be detected. For large enough tunnel coupling,
the negativity shows a pronounced maximum at an intermediate 
interaction strength within the Wigner molecule regime. 
\end{abstract}

Semiconductor few-electron quantum dots (QDs) 
continue to attract a lot of interest, as  
it has now become possible to experimentally control both
the electronic spin and charge in a condensed-phase 
environment in an unprecedented manner.
In particular, double quantum dots (DQDs) can be 
fabricated in a well-controlled fashion in
high-quality semiconductor devices, and  
are currently under intense study 
 \cite{ddot1,elzermann,ddot2,ddot3,ddot4,ddot5,ddot6,ddot7,ddot8}. 
Vertical or lateral tunnel-coupled DQDs are among the most promising 
candidates for realizing spin or charge qubits in a
quantum information processor \cite{loss1,Loss,LossMacDonald}. 
Their main advantages are scalability,  good control of 
physical properties via tunable external (sidegate-) voltages or
 magnetic fields,  and spatial separation of the individual
QDs (allowing to perform one- or two-qubit operations). 
Recent progress has been very swift, and present-day experiments 
are performed on DQDs containing just one or two electrons.

We present an exact diagonalization study of
ground-state entanglement in a vertical tunnel-coupled DQD containing 
two interacting electrons.  In general, entanglement 
provides a crucial  resource for quantum computing,
making certain tasks faster or more secure \cite{nielsen}. While 
coherent single-electron dynamics has been successfully
realized in DQDs, see, e.g., Refs.~\cite{ddot7,ddot8}, 
systematic studies of the two-electron dynamics and 
of entanglement are only now coming into reach \cite{ddot3}. 
In view of these developments, it seems timely to provide
theoretical predictions for two-electron charge
entanglement in DQDs, including both the effects of 
electron-electron interactions and of spin-orbit (SO) 
couplings.  Entanglement of two `charge qubits', which here
arise because an electron may reside in the 
upper or the lower dot, in such a bipartite mixed state
can, for instance, be determined by the Peres-Horodecki measure 
(the `negativity' $N$) \cite{ph,Thorwart}
in a mathematically sufficient and necessary way.  
Other entanglement measures exist \cite{Bruss}, for instance, 
the commonly used concurrence $C$ \cite{Wootters}, which is 
mathematically equivalent. 
It obeys the inequality $C\geq N$ \cite{Equiv,foot0},
although we find only small deviations between $C$ and $N$ (of order
$10^{-3}$ in our case).  
For sufficiently weak magnetic field, the two-electron system
is always in a spin singlet state,
and therefore the electrons have opposite spin
projections. Since both charge qubits physically occupy the same DQD, 
this is essential to make entanglement useful:
spin projection allows to distinguish different charge qubits, 
and thereby charge entanglement detection and/or exploitation
becomes possible.
We outline a concrete proposal below. 

Starting from a separable (non-entangled) state in the non-interacting limit,  
entanglement is enhanced with increasing interaction strength between 
the two charge qubits. 
For small tunnel coupling $\Delta$ between the QDs, only very
weak interactions are necessary to entangle the qubits, see Eq.~(\ref{lc})
below, allowing for efficient entanglement generation. 
Interestingly, for sufficiently large $\Delta$, 
$N$ exhibits a maximum at an intermediate interaction strength,
with suppressed entanglement for both weaker and stronger 
interactions.  The `optimal' interaction strength corresponds to
a Brueckner parameter $r_s\gtrsim 2$, where
Wigner molecule formation is expected \cite{reimann,Weiss}.
Our results imply that entanglement of two charge qubits in DQDs 
can be generated and controlled by gate electrodes, as these
affect the lateral size of the QDs and hence the effective 
interaction strength.

We consider two identical QDs with a two-dimensional (2D) 
parabolic confinement (in the $x$-$y$ plane) of frequency $\omega_0$.
In the perpendicular ($z$) direction, we take
a very steep confinement such that only the lowest state $|L\rangle$
($|U\rangle$) will be occupied in the lower (upper) QD. 
For a given electron, this two-level system defines the 
charge qubit of interest here.
We assume a distance $d$ between the QD centers along the $z$ axis,
leading to a tunneling amplitude $\Delta$ coupling the QDs.  
The in-plane coordinates of the two electrons are 
${\bf r}_{1,2}=(x_{1,2},y_{1,2})$, while the 
$z$-coordinate is described by the eigenvalue $\tau_i=\pm 1$
of a pseudo-spin matrix $\tau^z$, where the
 Pauli matrices $\tau^{x,y,z}$ act in the charge qubit ($|L,U\rangle$)
space. With dielectric constant $\kappa$ of the substrate,  
effective mass $m^\ast$, and dimensionless Rashba SO coupling
$\alpha_R$,
\begin{eqnarray} \label{modelham}
H&=&\sum_{i=1,2}\Bigl [\frac{{\bf p}^2_i}{2m^*}
+\frac{m^*\omega_0^2{\bf r}^2_i}{2}-\frac{\Delta}{2}\tau_i^x 
+ \omega_0 l_0 \alpha_R  \\
&\times& \nonumber  (p_{x,i}\sigma_i^y-p_{y,i}\sigma_i^x)\Bigr]
+\frac{e^2/\kappa}
{\sqrt{({\bf r}_1-{\bf r}_2)^2+d^2\delta_{\tau_1,-\tau_2}}}. 
\end{eqnarray}
Since the Dresselhaus term or other SO contributions 
due to the confinement are typically weaker 
and not tunable by gate voltages, we only keep the Rashba term;
Pauli matrices $\sigma^{x,y,z}$ act in spin space.
We use $l_0=\sqrt{\hbar/m^*\omega_0}$, the
dimensionless interaction parameter $\lambda=e^2/(\hbar \kappa l_0\omega_0)$
related to $r_s$  \cite{reimann}, and put $\Delta=\hbar \omega_0 \exp(-d/l_0)$,
consistent with the 
$\delta$-function confinement along the $z$-direction.

With integer radial ($n\geq 0$) and angular momentum ($M$) 
quantum numbers, the $\alpha_R=\lambda=0$ 
spin-degenerate single-particle eigenstates to energy   
$E_{nM}^{\pm}=(2n+|M|+1)\hbar \omega_0 \mp \Delta/2$ are 
\begin{eqnarray*}
\psi_{nM}^{\pm}(r,\varphi)&=&\left(\frac{n!}{2\pi l^2_0 (n+|M|)!}\right)^{1/2}
 \times (|L\rangle\pm|U\rangle) \\
&\times& e^{iM\varphi} (r/l_0)^{|M|} 
\ e^{-r^2/2l_0^2} \ \mathcal{L}_n^{|M|}(r^2/l_0^2) 
\end{eqnarray*}
with associated Laguerre polynomials $\mathcal{L}_n^{m}$. 
The Hamiltonian (\ref{modelham}) is represented in the 
product basis 
$\{|UU\rangle,|UL\rangle,|LU\rangle,|LL\rangle\}\otimes\{|\uparrow\uparrow
\rangle,|\uparrow\downarrow\rangle,|\downarrow\uparrow\rangle,
|\downarrow\downarrow\rangle\} \otimes(n_c,M_c)\otimes(n_r,M_r),$
where we use center-of-mass (COM) and relative quantum numbers 
$(n_{c,r},M_{c,r})$, respectively. 
For $\alpha_R=0$, a singlet spin state $S=0$ is realized for arbitrary 
$\lambda$ \cite{ddot1}, and then we restrict the representation to this sector. 
Moreover, the COM part completely decouples in that case. 
For $\alpha_R\ne 0$, however, this decoupling does not hold anymore, 
no assumptions about the spin sector are allowed, and the 
full problem has to be diagonalized.  (Complete spin
entanglement was found for all $\alpha_R$ studied below.) 
In the exact diagonalization scheme,
we impose an UV cutoff in $(n,M)$ space and increase it until 
convergence is achieved.  Given the representation of $H$,
the antisymmetric two-electron state $|\Psi_0\rangle$ 
can be computed, and hence the $T=0$ density 
matrix $|\Psi_0\rangle \langle\Psi_0|$.  To study 
charge qubit entanglement,
we form the $4\times 4$ reduced density matrix $\rho$
by tracing out the spin and $(n_{c,r},M_{c,r})$ 
degrees of freedom.  The Peres-Horodecki measure
(i.e. the {\sl negativity}) then follows as $N=2|\zeta_{\rm min}|$ \cite{ph},
where $\zeta_{\rm min}$ is the smallest eigenvalue of the 
partially transposed $\rho^{T_2}$ associated with the
mixed state described by $\rho$. The partially transposed of the reduced
two-particle density matrix $\rho_{\tau_1\tau_2;
\tau_1'\tau_2'}$ (where $\tau=U,L$) is defined as
$\rho^{T_2}_{\tau_1\tau_2;\tau_1'\tau_2'}=
\rho_{\tau_1'\tau_2;\tau_1\tau_2'}$, i.e. one has to transpose
with respect to one of the two electrons only.  
For mixed bipartite systems, the negativity 
provides a mathematically rigorous description of entanglement
 \cite{ph,Thorwart,Wootters,Bruss,Equiv}.
For a separable (non-entangled) state, $N=0$, 
whereas for maximally entangled states, $N=1$.

\begin{figure}
\vspace{0.2cm}
\includegraphics[width=70mm]{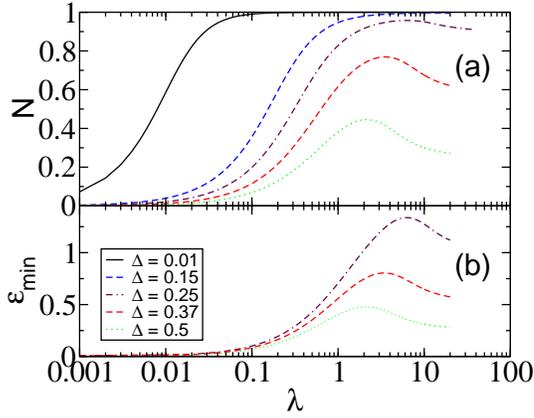}
\caption{ \label{CapvarD} (Color online)
(a) Negativity $N$ versus interaction strength $\lambda$.  
Results are
shown for several $\Delta$ (in units of 
$\hbar \omega_0$), with $\alpha_R=0$. 
(b) Corresponding energy scale $\varepsilon_{\rm min}$ (in units
of $\hbar\omega_0$), 
see Eq.~(\ref{emin}), as a function of $\lambda$.  (Data for
$\Delta/\hbar\omega_0\le 0.15$ are not shown, 
since $\varepsilon_{\rm min}$ 
has no maximum.) Note the semi-logarithmic scale.}
\end{figure}

Let us then discuss results, starting with $\alpha_R=0$.
In the absence of interactions, $\lambda=0$, both electrons occupy
the same pseudo-spin state $(|L\rangle+|U\rangle)/\sqrt{2}$ with
different spin. The two-electron charge state thus factorizes, and $N=0$.   
In the presence of interactions, however,
we find a finite negativity $N$, see Fig.~\ref{CapvarD},
indicating stronger entanglement with increasing $\lambda$
for all $\Delta$ and $0<\lambda\lesssim 1$.  This 
can be rationalized in terms of a simple energy scale argument. 
For small $\lambda$, only the transverse
ground state is occupied ($n_r=M_r=0$).  
Interactions then tend to localize electrons in the 
$|L\rangle, |U\rangle$ states, and
antisymmetry of the total (spin-singlet) wavefunction  
implies a symmetric two-electron pseudo-spin state, 
$(|L_1 U_2\rangle +|U_1 L_2\rangle)/\sqrt{2}$.
 Tunneling opposes this localization tendency, and a balance is
reached for some value $\lambda_c$. 
We take $N(\lambda_c)=1/2$ to define $\lambda_c$,
which sets the  crossover scale from weak to 
strong entanglement.  An estimate for $\lambda_c$ follows by 
equating the two relevant energy scales, 
\begin{equation} \label{lc}
\lambda_c = (\Delta/\hbar\omega_0) \ |\ln(\Delta/\hbar\omega_0)|.
\end{equation}
Detailed comparison shows that Eq.~(\ref{lc}) is in excellent agreement with 
the values following from the numerical results together with the
definition $N(\lambda_c)=1/2$.
For small $\Delta$, one reaches a maximally entangled Bell state ($N=1$)
already for very weak interactions, see Fig.~\ref{CapvarD}, which 
demonstrates that charge entanglement can be efficiently generated.

The data shown in Fig.\ \ref{CapvarD}(a) reveal an interesting  
feature for sufficiently large tunnel couplings,
$\Delta/\hbar\omega_0 \gtrsim 0.25$, namely a 
pronounced {\sl entanglement maximum} at an intermediate interaction
strength $\lambda_m>\lambda_c$.  
Note that tunable $\Delta$ reaching a few meV have been reported in
recent DQD experiments \cite{ddot2,ddot7}. Since typically
$\hbar\omega_0\lesssim 10$~meV, our parameter
range for $\Delta$ is accessible. 
While for $\Delta/\hbar \omega_0 \lesssim 0.2$,  $N= 1$
for  $\lambda\gg\lambda_c$, in the strong-tunneling case, entanglement
decreases again for $\lambda>\lambda_m$.
The maximum shifts to lower $\lambda_m$
when tunneling is increased.  To give concrete numbers, 
$\lambda_m=3.53$ for $\Delta/\hbar\omega_0=0.37$, while 
$\lambda_m=2.13$ at $\Delta/\hbar\omega_0=0.5$, 
corresponding to $r_s\approx 4.8$ and $2.4$, respectively.   
These interactions indicate that one has reached 
the incipient Wigner molecule regime \cite{Weiss,reimann}, where
interactions become strong enough to induce Wigner crystallization
in this finite-size system.  The $r_s$ values necessary for observing
the maximum in $N(\lambda)$ are large but within reach of present
experiments \cite{ddot1}.

\begin{figure}
\vspace{0.2cm}
\includegraphics[width=70mm]{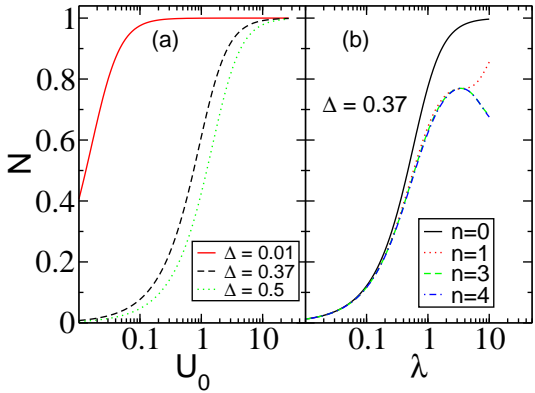}
\caption{ \label{SModel} (Color online)
Negativity versus interaction strength
for (a) the simple model $H_s$ in Eq.~(\ref{simpleham}), with several 
different $\Delta$ (in units of $\hbar \omega_0$), and (b) 
for the full  $H$ at $\Delta/\hbar\omega_0=0.37$,
with the number of oscillator states $n_r$ truncated to $n_r\leq n$
(for several $n$).  }
\end{figure}

In order to understand this behavior, it is crucial to 
address the role of higher transverse states in the individual
QDs.  Those states play no significant role for weak interactions,
but can be populated for strong interactions. 
To elucidate this, consider the following simple DQD model with
two spin-degenerate levels corresponding to the lower/upper dot,
\begin{equation}
H_s=U_0\sum_{j=L,U} n_{\uparrow,j} n_{\downarrow,j} + V_0 n_L n_U -
\Delta\sum_\sigma (d_{\sigma,L}^{}d_{\sigma,U}^\dagger+ {\rm h.c.}),
\label{simpleham}
\end{equation}
where $d_{\sigma j}$ is the annihilation operator for
spin $\sigma=\uparrow,\downarrow$ in QD $j=L/U$, 
$n_{\sigma j}=d_{\sigma j}^\dagger d_{\sigma j}^{}$, and
$n_j=n_{\uparrow,j}+n_{\downarrow,j}$ with $n_L+n_U=2$.
The interaction between electrons on the same dot (on different dots) is 
$U_0$ $(V_0)$. 
The negativity can then be calculated in the same way as
before and is shown as a function of $U_0$ in Fig.~\ref{SModel},
with the order-of-magnitude estimate $V_0=U_0/(d/l_0)$.  
These results indicate that for any tunneling amplitude $\Delta$, 
entanglement becomes stronger with increasing interactions.
The same conclusion is reached by using more sophisticated
estimates for $U_0, V_0$, starting from Eq.~(\ref{modelham}) and
projecting the full Hamiltonian to the transverse ground state.
We conclude that $H_s$ is not able to 
reproduce the suppression of entanglement at large $\lambda$ seen in 
Fig.~\ref{CapvarD}(a), and therefore higher transverse states of the QDs are
crucial in understanding this effect.  
In fact,  keeping only   
oscillator states with $n_r\leq n$ in the diagonalization of 
Eq.\ (\ref{modelham}), 
Fig.~\ref{SModel}(b) shows that with increasing $\lambda$ higher 
states become
more and more relevant. For $\Delta/\hbar\omega_0=0.37$, see
Fig.~\ref{SModel}(b), we recover the entanglement maximum once $n\geq 3$.
The relevance of these higher-lying transverse states 
for the suppression of entanglement also follows from an energetic
argument involving the competition of tunneling and 
Coulomb repulsion. For this, consider the energy difference 
\begin{equation}\label{emin}
\varepsilon_{\rm min}(\lambda, \Delta)=E_{\rm same}-E_{\rm diff} ,
\end{equation}
where the two electrons either reside on the same dot ($E_{\rm same}$) or
on different dots ($E_{\rm diff}$).  For $\lambda=0$, we have
$\varepsilon_{\rm min}=0$, corresponding to degenerate pseudo-spin states
and therefore to $N=0$.  In general, we can then expect that a correlation
between $N$ and $\varepsilon_{\rm min}$ exists, with large (small) 
$\varepsilon_{\rm min}$ corresponding to large (small) $N$. 
This correlation is natural \cite{Loss}, since the
entangled state corresponds to the electrons occupying
different dots, leading to the energy $E_{\rm diff}$, while
entanglement is destroyed once both electrons reside on the same dot.
Diagonalizing $H$ separately for these two cases, 
the numerical result for $\varepsilon_{\rm min}$ is shown for various $\Delta$
as a function of $\lambda$ in Fig.~\ref{CapvarD}(b). 
The energy difference $\varepsilon_{\rm min}$ grows with increasing
$\lambda$ for  $\Delta/\hbar \omega_0\lesssim 0.2$, but has a maximum for
 $\Delta/\hbar \omega_0\gtrsim 0.25$. 
Remarkably, the behavior of $\varepsilon_{\rm min}$ 
perfectly reflects what we obtain for the negativity.
The decrease in $\varepsilon_{\rm min}$ with large 
$\lambda$ does not occur
for $H_s$ (data not shown), which again points to the importance
of higher-lying transverse states. 
Intuitively, the interaction energy is minimized by pushing the two electrons 
to opposite sides in the $x$-$y$ plane, which corresponds to occupation of 
these states. 
By that mechanism, the Coulomb repulsion between electrons located
on different dots is efficiently reduced, and thereby entanglement
can be suppressed.  We mention that 
  for very large interactions,
a simple classical calculation (i.e., neglecting the kinetic energy) shows 
that 
$\varepsilon_{\rm min}$ reaches a constant value, 
$\varepsilon_{min}\propto d^2$.  This in turn implies by the above
 argumentation that the negativity does 
not vanish as $\lambda\to \infty$, but instead stays at a finite value.

\begin{figure}
\vspace{0.2cm}
\includegraphics[width=70mm]{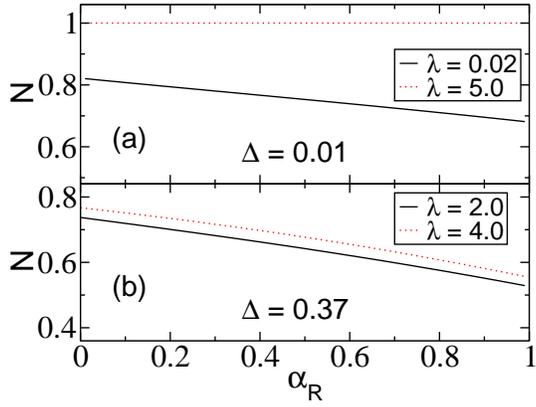}
\caption{ \label{alphaR} (Color online)
Negativity as function of the SO coupling $\alpha_R$ for several
$\lambda$, for (a) $\Delta/\hbar \omega_0=0.01$ and 
(b) $\Delta/\hbar \omega_0=0.37$.}
\end{figure}

Let us then briefly address the effects of finite 
SO couplings $\alpha_R$ on charge entanglement, see Fig.\ \ref{alphaR}.
In general, SO couplings tend to weakly decrease $N$.
For InAs dots, where SO couplings are expected to be quite strong,   
values in the regime $\alpha_R\lesssim 1$ were measured \cite{Nitta}. 
For such couplings, we find that $N$ decreases by at most $20\%$. 
The decrease is most pronounced for strong tunneling between the 
dots and/or strong electron-electron interaction.  
In the presence of various types of SO couplings, 
already for just one electron in the DQD
one may expect `hyperentanglement' \cite{walborn} of spin and charge.
However, keeping only the
Rashba coupling in Eq.~(\ref{modelham}) 
does not lead to a non-zero hyperentanglement negativity,
even when allowing for different SO couplings in the two dots.

\begin{figure}
\vspace{0.2cm}
\includegraphics[width=84mm]{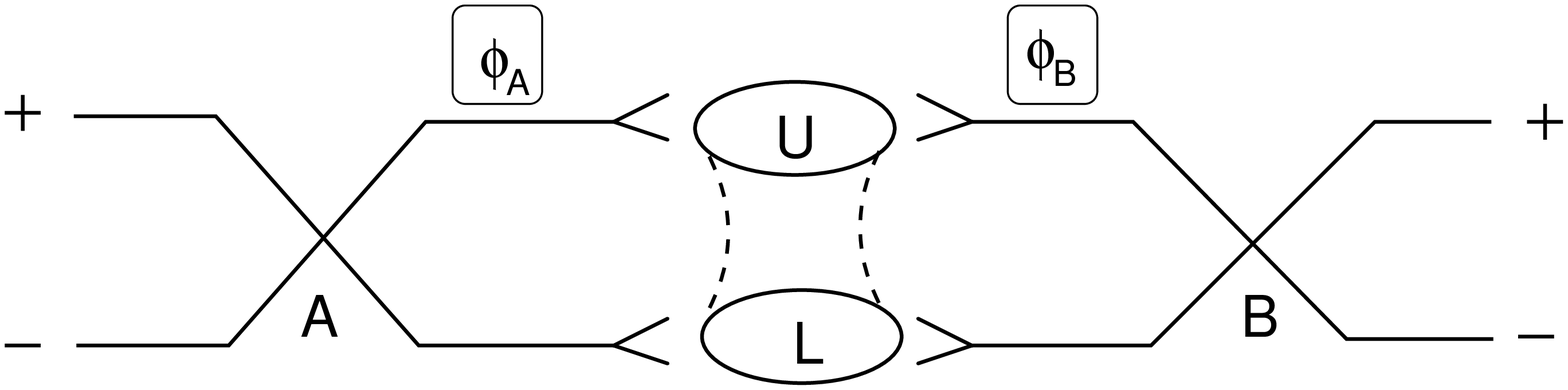}
\caption{ \label{fig4} 
Schematic setup to detect charge qubit entanglement adopted from Ref.~
\cite{Samuelsson06}. The upper and lower
dot in the DQD are indicated by $U$ and $L$, respectively. 
The DQD is connected by switchable barriers to two beam splitters 
$A$ and $B$, and two side gates induce  controlled phase shifts
$\phi_{A/B}$.  }
\end{figure}

When tuning entanglement by adiabatic manipulations such as
slow changes of gate voltages, some relaxation mechanism 
is implicitly needed for equilibration. However, we require relaxation to be
weak enough, for otherwise quantum coherence is affected or even destroyed.
Typical charge relaxation times in high-quality DQDs 
presently approach $1\mu$s \cite{ddot8}, 
and thus the coherence requirements should pose no major obstacle.
 Let us then discuss how the charge qubit entanglement produced in the DQD 
 can be detected and/or exploited by application of the
scheme suggested in Ref.~\cite{Samuelsson06}, see Fig.~\ref{fig4}. 
Two electronic beam splitters \cite{Henny99}
are attached together with two side gates in one input arm of each. 
The latter  allow to induce controlled phase
shifts $\phi_A$ and $\phi_B$ on the orbital states \cite{phaseshift}. 
The accessible observables are average
currents $I_{\nu}^\alpha$ and zero-frequency current correlators 
$S_{\nu,\mu}^{\alpha,\beta}$ with $\nu,\mu=A,B$ and $\alpha,\beta=\pm$, 
see Fig.~\ref{fig4}. Suppose now that the contacts to the beam splitters 
can be individually addressed by adiabatic gate voltage 
pulses switching them from `closed' to `open'. Two electrons
 are then emitted from the double dot and enter
the detection region. The two-particle reduced density matrix
(and hence $N$) then follows from
16 current correlation and 8 average current measurements
\cite{Samuelsson06}. 

To conclude, we have studied charge qubit entanglement in 
double quantum dots.  Entanglement can be created and altered
by adiabatic changes of electrostatic potentials.
While we have specifically discussed  vertical double dots, our
general conclusions also apply to lateral double dots or 
carbon-nanotube based dots.  The case of more than 
two interacting electrons remains as an interesting open challenge, 
where the mathematical foundations for 
entanglement measures are less clear.  

We thank D. Bru{\ss}, T. Heinzel, and H. Kampermann for 
discussions.  This work was supported by the ESF network INSTANS.

\begin{thebibliography}{0}

\bibitem{ddot1} 
\Name{van der Wiel W. G., De Franceschi S., Elzerman J. M., Fujisawa T.,
 Tarucha S., \and  Kouwenhoven L.}
\REVIEW{Rev. Mod. Phys.}{75}{2003}{1}.

\bibitem{elzermann}
\Name{Elzerman J. {\sl et al.}} 
\REVIEW{Phys. Rev. B}{67}{2003}{161308(R)}.

\bibitem{ddot2}
\Name{Hatano T., Stopa M., \and Tarucha S.} 
\REVIEW{Science}{309}{2005}{268}.

\bibitem{ddot3}
\Name{Petta J. R. {\sl et al.}}
\REVIEW{Science}{309}{2005}{2180}.

\bibitem{ddot4}
\Name{Johnson A. C. {\sl et al.}}
\REVIEW{Nature}{435}{2005}{925}.

\bibitem{ddot5}
\Name{Koppens F. H. L. {\sl et al.}}
\REVIEW{Science}{309}{2005}{1346}.

\bibitem{ddot6}
\Name{Krenner H. J.{\sl et al.}}
\REVIEW{Phys. Rev. Lett.}{94}{2005}{057402}.

\bibitem{ddot7}
\Name{H{\"u}ttel A. K., Ludwig S., Lorenz H., Eberl K.,\and Kotthaus J.}
\REVIEW{Phys. Rev. B}{72}{2005}{081310(R)}.

\bibitem{ddot8}
\Name{Gorman J., Hasko D. G., \and  Williams D. A.} 
\REVIEW{Phys. Rev. Lett.}{95}{2005}{090502}.

\bibitem{loss1}
\Name{Loss D. \and  DiVincenzo D. P.}
\REVIEW{Phys. Rev. A}{57}{1998}{120}.

\bibitem{Loss} 
\Name{ Burkard G., Seelig G., \and Loss D.} 
\REVIEW{Phys. Rev. B}{62}{2581}{2000}; 
\Name{Golovach V. N. \and  Loss D.} 
\REVIEW{Phys. Rev. B}{69}{2004}{245327}.

\bibitem{LossMacDonald} 
\Name{ Schliemann J.,  Loss D., \and  MacDonald A. H.} 
\REVIEW{Phys. Rev. B}{63}{2001}{085311}.

\bibitem{nielsen}
\Name{Nielsen M. A. \and  Chuang I. L.} 
\Book{Quantum Computation and Quantum Information}\Publ{Cambridge
University Press} \Year{2000}.

\bibitem{ph}
\Name{Peres A}.
\REVIEW{Phys. Rev. Lett.}{77}{1996}{1413};
\Name{ Horodecki M., Horodecki P., \and  Horodecki R.} 
\REVIEW{Phys. Lett. A}{223}{1996}{1}. 

\bibitem{Thorwart}
\Name{ Thorwart M. \and H\"anggi P.}
\REVIEW{Phys. Rev. A}{65}{2002}{012309}. 


\bibitem{Bruss}
\Name{Bru{\ss} D.}
\REVIEW{J. Math. Phys.}{43}{2002}{4237}.

\bibitem{Wootters}
\Name{ Wootters W. K.}
\REVIEW{Phys. Rev. Lett}{80}{1998}{2245}.

\bibitem{Equiv}
\Name{ Verstraete F.,  Audenaert K.,  Dehaene J., \and  De Moor B.} 
\REVIEW{J. Phys. A}{34}{2001}{10327}.

\bibitem{foot0}
The measure introduced in Ref.~\cite{LossMacDonald} here equals $C$.  

\bibitem{reimann} 
\Name{Reimann S. M. \and  Manninen M.} 
\REVIEW{Rev. Mod. Phys.}{74}{2002}{1283}.

\bibitem{Weiss}
\Name{Weiss S. \and  Egger R.}
\REVIEW{Phys. Rev. B}{72}{2005}{245301}.

\bibitem{Nitta}
\Name{Nitta J. {\sl et al.}}
\REVIEW{Phys. Rev. Lett.}{78}{1997}{1335};
\Name{Koga T. {\sl et al.}}
\REVIEW{Phys. Rev. Lett.}{89}{2002}{046801}.

\bibitem{walborn}
\Name{Walborn S. P.,  Padua S., \and  Monken C.H.}
\REVIEW{Phys. Rev. A}{68}{2003}{042313}.

\bibitem{Samuelsson06}
\Name{ Samuelsson P. \and  B\"uttiker M.}
\REVIEW{Phys. Rev. B}{73}{2006}{041305(R)}.

\bibitem{Henny99}
\Name{Henny M. {\sl et al.}}
\REVIEW{Science}{284}{1999}{296}; 
\Name{Oliver W. D., Kim J., Liu R., \and  Yamamoto Y.}
\REVIEW{Science}{284}{1999}{299}.

\bibitem{phaseshift}
\Name{Ji Y. {\sl et al.}}
\REVIEW{Nature}{422}{2003}{415}. 

\end{thebibliography}
\end{document}